\documentclass[fleqn,twoside]{article}
\usepackage[headings]{espcrc2}

\readRCS
$Id: espcrc2.tex,v 1.2 2004/02/24 11:22:11 spepping Exp $
\usepackage{graphicx}
\usepackage[figuresright]{rotating}

\newcommand{\AmS}{{\protect\the\textfont2
  A\kern-.1667em\lower.5ex\hbox{M}\kern-.125emS}}
\newcommand{\BlackHat}{{\sc BlackHat}}
\newcommand{\Res}{\mathop{\rm Res}}
\newcommand{\tree}{{\rm tree}}
\newcommand{\e}{\epsilon}

\newcommand{\Bmp}[1]{\langle #1\rangle}

\newcommand{\Kf}[1]{\widetilde K_{#1}}
\newcommand{\Kfmu}[1]{\widetilde K^{\mu}_{#1}}

\title{\vspace{-5cm}\rule{1cm}{0cm}\hfill {\small\rm MIT-CTP 3963\\\hfill \rm Saclay-IPhT-T08/117 \\\hfill \rm SLAC-PUB-13295\\\hfill    \rm UCLA/08/TEP/21\\}\rule{0cm}{0cm}\vspace{2cm}\\One-Loop Calculations with BlackHat}

\author{C.~F.~Berger\address{Center for Theoretical
Physics, Massachusetts Institute of Technology,
   Cambridge, MA 02139, USA},
Z.~Bern\address[UCLA]{Department of Physics and Astronomy, UCLA, 
   Los Angeles, CA 90095-1547, USA},
L.~J.~Dixon\address[SLAC]{Stanford Linear Accelerator Center, 
   Stanford University, Stanford, CA 94309, USA},
F.~Febres Cordero\addressmark[UCLA],
D.~Forde\addressmark[SLAC],
H. Ita\addressmark[UCLA],
D.~A.~Kosower\address{Institut de Physique Th\'eorique, CEA--Saclay,
          F--91191 Gif-sur-Yvette cedex, France}\thanks{Presenter},
D.~Ma\^{\i}tre\addressmark[SLAC]\thanks{Presenter}}
       
\runtitle{One-loop Calculations with BlackHat}
\runauthor{D. Ma\^{\i}tre et al.{}}

\begin{document}

\begin{abstract}
  We describe \BlackHat{}, an automated C++
  program for calculating one-loop amplitudes, and the techniques used
  in its construction.  These include
  the unitarity method and on-shell recursion. The other ingredients
  are compact analytic formulae for tree amplitudes for 
  four-dimensional helicity states.  The program
  computes amplitudes numerically, using analytic formul\ae{} only for
  the tree amplitudes, the starting point for the recursion,
  and the loop integrals. We make
  use of recently developed on-shell methods for evaluating coefficients
  of loop integrals, in particular a discrete Fourier projection as a means 
  of improving numerical stability.  
  We illustrate the good numerical stability of this approach by
  computing six-, seven- and eight-gluon amplitudes in QCD
  and comparing against known analytic results.
\vspace{1pc}
\end{abstract}

\maketitle

\section{Introduction}
\label{Introduction}

Quantitatively reliable predictions for background processes will play
an important role in ferreting out signals of new physics in experiments
at the upcoming Large Hadron Collider (LHC).  New physics beyond the
Standard Model is expected to emerge in these TeV-scale experiments.
Known-physics backgrounds from electroweak, QCD, and mixed processes
will also contribute events that may overwhelm or mimic new-physics signals.
Uncovering and understanding the new-physics signals will require use
of elaborate kinematic requirements (such as several identified jets,
cuts on missing transverse energy, etc.) and reliable knowledge of
background processes.

Leading-order calculations in QCD suffer from large uncertainties
and therefore do not suffice to furnish the required
quantitative knowledge of backgrounds.  Next-to-leading order calculations
are required~\cite{NLMLesHouches}.  Indeed, for a few signal, background,
or calibration processes (Higgs-boson production, top production,
and distributions
associated with production of single electroweak vector bosons), precision
calculations at next-to-next-to-leading order (NNLO) are needed.
For other processes, NLO will likely suffice until the LHC enters
its precision-measurement era.  There are a large number of processes
that ought to be computed, however, and these include many processes
with many final-state jets, corresponding to large final-state parton
multiplicity.  

NLO predictions can be provided by parton-level Monte Carlo programs,
or by parton showers such as MC@NLO~\cite{MC@NLO}.  Both types of
program require the computation of virtual and real-emission amplitudes.
The computation of the latter relies on well-understood 
technologies~\cite{TreeLevelApproaches,TreeLevelPrograms}.  
Parton-level programs also require the isolation
of infrared singularities in the integrated real-emission contributions,
and a means of canceling them systematically against the virtual-correction
singularities.  These technologies~\cite{Slicing,Subtraction} 
are also well understood and recently authors have begun to automate
them~\cite{AutomatedSubtraction}.  The infrared-divergent parts of
the one-loop virtual corrections are also 
well-understood~\cite{Slicing,KST}.  The remaining, infrared-finite
parts of these one-loop amplitudes have been the difficult part of
computations with three final-state objects, and have been the primary
bottleneck to computations of
processes with four or more final-state objects.  \BlackHat{} is
one of a new generation of numerical 
approaches that aims to break this bottleneck.
Other numerical efforts along similar lines are described in 
refs.~\cite{CutTools,EGK,GKM}.  

The traditional approach to one-loop computations uses Feynman diagrams.
With an increasing number of external particles, however, the computational
complexity of such computations with traditional methods grows factorially.
This complexity cannot be tamed by use of technologies such
as the spinor-helicity formalism~\cite{SpinorHelicity} alone.
Recent years have witnessed a ferment of development, 
based on the analytic properties of unitarity and
factorization that any amplitude must satisfy,
of new approaches to overcoming these computational 
difficulties~\cite{OtherApproaches,AguilaPittau,Binoth,Denner,EGZ,XYZ},
including on-shell methods~\cite{UnitarityMethod,Zqqgg,%
DDimUnitarity,BCFUnitarity,BCFRecursion,BCFW,%
DoublePole,Bootstrap,Genhel,OtherBootstrap,RecentDDim,%
OPP,BFMassive,OnShellReview,Forde,OPP2,MOPP,BOPP}.  
These methods
are efficient, and feature only mild growth in required computer time
with increasing number of external particles.  They effectively
reduce loop calculations to tree-like ones, simplifying them intrinsically
and further allowing use of efficient algorithms for the
tree-amplitude ingredients.


\section{The On-Shell Approach}

\BlackHat{} is built using an on-shell approach, the
unitarity method~\cite{UnitarityMethod,Fusing} with multiple
cuts~\cite{Zqqgg} ({\it a.k.a.\/} generalized unitarity), along with
significant 
refinements~\cite{BCFUnitarity,OPP,Forde} exploiting complex momenta. 
The cuts replace two or more propagators by delta functions, thus putting the corresponding momenta on shell.  They thereby isolate terms in the amplitude with distinct analytic structure, allowing them to be computed independently.

We employ a predominantly numerical approach in \BlackHat{}.  The
loop integrals are evaluated from analytic formul\ae{}, likewise
their contributions
to residues at spurious singularities (see section \ref{sec:rational}); and analytic
formul\ae{} may be used to speed up evaluation of tree amplitudes.
Otherwise, the code is numerical, and in particular everything that
corresponds to algebra in a symbolic calculation is done numerically.
This makes
it easier to design a general-purpose code, as distinct from the bespoke
analytic process-by-process calculations that have been done to date.  
It is also important, however, to obtain an efficient code.  As the
example of the Berends--Giele recursion relations shows~\cite{BGComplexity},
evaluating them recursively with caching of intermediate currents,
a numerical approach can make it much more practical to eliminate
repeated evaluation of common subexpressions than an analytic approach.
Indeed, it seems likely that only a numerical approach can meet
the goal of a polynomial-time algorithm for the evaluation of each helicity amplitude at one loop.

We begin by separating the amplitude into cut-containing and rational parts,
\begin{equation}
A_n = C_n + R_n \,,
\label{CutRational}
\end{equation}
where the former contain all (poly)logarithms, $\pi^2$ terms, and the
finite constant in the scalar bubble. The cut-containing part of massless dimensionally-regulated
amplitudes with four-dimensional external momenta may be written in
a basis of scalar 
integrals~\cite{IntegralsExplicit,IntegralReductions,BDKIntegrals,%
OtherPentagons,DuplancicNizic},
\begin{equation}
C_n = \sum_i d_i I_4^i + \sum_i c_i I_3^i + \sum_i b_i I_2^i \,,
\label{IntegralBasis}
\end{equation}
The integrals $I_{2,3,4}$ are respectively bubble, triangle, and box
integrals.

We compute the coefficients $b_i$, $c_i$, and $d_i$ numerically using
the unitarity method, with four-dimensional loop momenta.  The tree
amplitudes that feed into the computation may be evaluated efficiently
using spinorial methods.  We compute the rational parts using
loop-level on-shell recursion.

\section{Cut Parts}

The most straightforward coefficients to obtain are those of the box
integrals.  Imposing four cut conditions on a one-loop integrand, and
then setting $\e=0$, freezes it completely.  Furthermore, this isolates
the coefficient of a single box integral uniquely.  The coefficient is then
given~\cite{BCFUnitarity} in terms of the product of
four tree amplitudes at the corners of the box,
\begin{eqnarray}
d_i &=& {1\over 2} \sum_{\sigma = \pm} d_i^\sigma\nonumber\\
d_i^\sigma&=& A^\tree_{(1)} A^\tree_{(2)} A^\tree_{(3)} A^\tree_{(4)} 
\Bigr|_{l_i = l_i^{(\sigma)}} \,,
\label{QuadCutSolution}
\end{eqnarray}
where the cut loop momenta $l_i^{(\pm)}$ are the two solutions to the quadruple-cut on-shell conditions, labeled by $\sigma=\pm$.

In the case of triangle coefficients, we can impose three cut conditions.
Here, the cut conditions no longer freeze
the integrand completely; one degree of freedom is left over.  There
are different ways to parametrize this degree of 
freedom~\cite{AguilaPittau,OPP}; we use
the variant proposed by Forde~\cite{Forde},
\begin{equation}
l^{\mu}(t) = \Kfmu1 +
\Kfmu3 +
\frac{t}{2}\Bmp{\Kf1|\gamma^{\mu}|\Kf3}
+\frac{
\Bmp{\Kf3|\gamma^{\mu}|\Kf1}}{2t}
,
\label{TripleCutMomentum}
\end{equation}
where $K_{1,3}$ are the two (possibly massive) external momenta separated
by the cut propagator $l$;
$\gamma = \gamma_{\pm} = - K_1\cdot K_3 \pm\sqrt{\Delta}$;
$S_i = K_i^2$, and
\begin{equation}
\Kfmu1\!=\!\hat\alpha\frac{\gamma K_1^{\mu}\!\!+\!S_1 K^{\mu}_3}{\gamma^2-S_1S_3}\,,
\hskip 1 mm
\Kfmu3\!=\!-\hat\alpha'\frac{\gamma K_3^{\mu}\!\!+\!S_3 K_1^{\mu}}{\gamma^2-S_1S_3} \,,
\label{eq:def_gamma_pm}
\end{equation} 
\begin{equation}
\hat\alpha = \frac{\gamma S_3(S_1-\gamma)}{S_1S_3-\gamma^2}\,,\quad
\hat\alpha' = \frac{\gamma S_1(S_3-\gamma)}{S_1S_3-\gamma^2}\,.
\end{equation}

Once the remaining degree of freedom
is parametrized by $t$, the integrand has the following form,
\begin{eqnarray}
\lefteqn{A^\tree_{(1)} A^\tree_{(2)} A^\tree_{(3)} \Bigr|_{l_i = l_i(t)}= }&&\nonumber\\ 
&&\frac{c_{-3}}{t^3}
+\frac{c_{-2}}{t^2}
+\frac{c_{-1}}{t}
+c_0 + c_1 t + c_2 t^2 + c_3 t^3 \nonumber\\
&&
+ \sum_{{\rm poles}} \frac{d_i^\sigma}{\xi_i^\sigma (t-t_i^\sigma)}\,.
\end{eqnarray}
The sum over poles corresponds to the
various box contributions sharing the same triple cuts.  We
follow Ossola, Papadopoulos, and Pittau (OPP) \cite{OPP} in subtracting these
contributions from the integrand, leaving behind seven independent
coefficients.  Forde's parametrization isolates the desired 
coefficient $c_0$ by virtue of its analytic property --- it is
the constant as $t\rightarrow \infty$, or equivalently it can be extracted
as the residue of the integrand divided by $t$,
\begin{equation}
c_0 = {1 \over 2\pi i} \oint {d t \over t} \,  T_3(t) \,.
\end{equation}
We evaluate this contour integral using a discrete Fourier projection,
\begin{equation}
c_0 = {1\over 2p+1} \sum_{j = -p}^{p} 
                        T_3\Bigl(t_0 e^{2 \pi i j/(2p + 1)}\Bigr)\,,
\end{equation}
where $t_0$ is an arbitrary complex number.  The projection avoids the
potentially-unstable matrix inversion that could arise from simply
inverting a system of equations to solve for the seven coefficients
$c_{-3},\ldots,c_3$.  (The other coefficients are needed in order to
subtract triangle coefficients when computing in their turn the bubble
ones, and can also be obtained by a Fourier projection.)  The
subtraction of the box coefficients makes the discrete Fourier
projection {\it exact\/} and also allows for the flexibility in the
choice of $t_0$.  It thereby improves the numerical stability of the
calculation.

The bubble coefficients can be computed following the same approach,
subtracting both box and triangle contributions, and using a two-dimensional
discrete projection.  We refer the reader to ref.~\cite{BlackHatI}
for more details.

\begin{figure*}
\begin{center}
\includegraphics[width=.97\textwidth,clip]{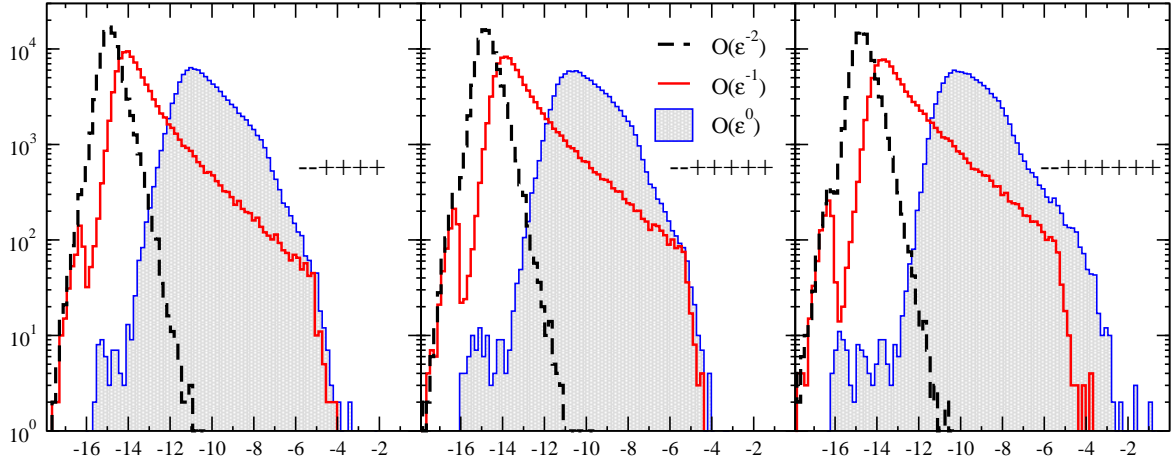}
\caption{The distribution of the logarithm of the relative error 
over 100,000 phase-space points for the MHV amplitudes
$A_6(1^-,2^-,3^+,4^+,5^+,6^+)$, $A_7(1^-,2^-,3^+,4^+,5^+,6^+,7^+)$ and
$A_8(1^-,2^-,3^+,4^+,5^+,6^+,7^+,8^+)$. The dashed (black) curve 
in each histogram gives the relative error 
for the $1/\e^2$ part,  the solid (red) curve gives the $1/\e$ 
singularity, and the shaded (blue) distribution gives 
the finite $\e^0$ component of the corresponding helicity amplitude.}
\label{fig:splitMHVs}
\end{center}
\end{figure*}

\section{Rational Parts}\label{sec:rational}
On-shell recursion relations for the rational terms may be derived by
considering deformations of the amplitude,
parametrized by a complex parameter $z$~\cite{BCFW}. These deformations shift two external momenta by $\pm z\cdot q$ where $q$ is a complex null four-vector, so as to preserve overall momentum conservation and leave all external momenta on shell. The recursion relation follows from evaluating the
contour integral
\begin{equation}
R^{{\rm large}\ z}_n={1\over 2\pi i} \oint_C dz \,{R_n(z)\over z} \,,
\end{equation}
where $C$ is a circle at infinity. If $R^{{\rm large}\ z}_n$ does not vanish for a given choice of deformation, it can be computed using an auxiliary recursion relation~\cite{Genhel}.
The rational terms may be computed using Cauchy's theorem, 
\begin{equation}
R_n(0) = R^{{\rm large}\ z}_n -\sum_{{\rm poles}\ \alpha} \Res_{z=z_\alpha}  
{R_n(z)\over z} \,.
\label{ResidueSum}
\end{equation}
The sum over the poles in the last term can be decomposed into two sets, the physical or spurious poles, depending on whether the pole is or is not present in the full deformed one-loop amplitude,  
\begin{equation}
R_n = R^D_n + R^S_n + R^{{\rm large}\ z}_n\,. 
\label{totrec}
\end{equation}

The contributions $R^D_n$ from the physical poles can be computed
using on-shell recursive diagrams~\cite{Bootstrap}.

In \BlackHat{}, the residues at the spurious poles are computed
using the cut parts.  Because the
spurious poles must cancel in the amplitude as a whole, we have
\begin{equation}
R^S_n =
 -\!\!\sum_{{\rm spur.}\atop{\rm poles}\ \beta} \Res_{z_\beta} {R_n(z)\over z}
= \!\!\sum_{{\rm spur.}\atop{\rm poles}\ \beta} 
  \Res_{z_\beta} {C_n(z)\over z} \,,
\end{equation}
where $C_n(z)$ is the cut part from eq.~(\ref{CutRational}), as deformed by
the on-shell deformation parametrized by $z$.

The spurious singularities arise from zeros of Gram determinants
implicitly appearing in the denominators of the integral coefficients
$b_i$, $c_i$, and $d_i$ of eq.~(\ref{IntegralBasis}). 
We evaluate the residues of these singularities by making a discrete approximation to a contour integral on a small circle around the pole.  At each complex value on the circle, we evaluate
the coefficients $b_i$, $c_i$, and $d_i$ numerically as described in
the previous section. Since the residues
are of course rational, and can only arise from expanding the dilogarithms
or logarithms in the integral functions, we do not directly evaluate the loop integrals 
numerically; rather, we first expand them analytically in the appropriate
neighborhood of the Gram-determinant singularity, and then evaluate
the rational expansion coefficients numerically.

We can control
the approximation by choosing the size of the circles around the spurious 
singularity and the number of points on the circle.  
(The current code evaluates the integrand at ten points
around the circle.)  This risks numerical instabilities if the circle
becomes too small.

In order to handle possible numerical instabilities, we test for them
dynamically, that is event by event and spurious pole by spurious
pole.  We test to see that the coefficient of the non-logarithmic
$1/\e$ singularity is reproduced correctly.  It is sufficient to test
the bubble coefficients, because they produce this singular term. Box and triangle coefficients are tested indirectly, because they are subtracted in order to
compute the bubble coefficients.  We also test to see that the
spurious singularities cancel in the sum over bubble coefficients.  If
either of these tests fail, we recompute the numerically unstable terms
of the amplitude at higher precision.  Because the fraction of points
failing the tests is small, this does not impose a significant time
penalty on our code.

\section{Results}

As an example, we used \BlackHat{} to compute
the one-loop $2\rightarrow4$-, $2\rightarrow5$-, and $2\rightarrow6$-gluon 
maximally helicity-violating (MHV)
amplitudes 
for $n_{\!f}=0$ at 100,000 phase-space points,
generated using a flat distribution.  (We impose the following
cuts: 
$E_T > 0.01 \sqrt{s}$, $\eta<3$, and $\Delta_R >0.4$.)
We compared the numerical results against those computed
using known analytic results.
The histogram in fig.~{\ref{fig:splitMHVs}} shows the results.  The horizontal axis
gives the logarithmic relative error,
\begin{equation}
\log_{10}\left(\frac{|A_n^{\rm num}-A_n^{\rm target}|}
                                {|A_n^{\rm target}|}\right)\,,
\end{equation}
for each of the $1/\e^2$, $1/\e$, and $\e^0$ parts of the one-loop
amplitude.
The
vertical axis in these plots shows the number of phase-space
points in a bin that agree with the target to a specified
relative precision.  The vertical scale is logarithmic, which
enhances the visibility
of the tail of the distribution, and thereby illustrates the good
numerical stability of the computation.

\section{Acknowledgments}
We thank Academic Technology Services at UCLA for computer support.  This
research was supported by the US Department of Energy under contracts
DE--FG03--91ER40662 and DE--AC02--76SF00515.  CFB's research was
supported in part by funds provided by the U.S. Department of Energy
(D.O.E.) under cooperative research agreement DE--FC02--94ER40818.  DAK's research is supported by the Agence Nationale
de la Recherce of France under grant ANR--05--BLAN--0073--01.
The work of DM was supported by the Swiss National Science Foundation
(SNF) under contract PBZH2--117028.

\end{document}